\documentclass[twocolumn]{aastex63}

\usepackage{xcolor}

\newcommand{\lam}{$\lambda$}

\renewcommand{\ion}[2]{#1\,{\sc #2}}

\received{June 1, 2019}
\revised{January 10, 2019}
\accepted{\today}
\submitjournal{ApJ}

\graphicspath{{./}{figures/}}

\begin{document}

\title{Fe\,VII emission lines in the wavelength range 193--197 \AA}

\author[0000-0001-9034-2925]{Peter R. Young}
\affiliation{NASA Goddard Space Flight Center, Greenbelt, MD 20771, USA}
\affiliation{Northumbria University, Newcastle upon Tyne, NE1 8ST, UK}

\author{Alexander N. Ryabtsev}
\affiliation{Russian Academy of Sciences, Institute of Spectroscopy, Troitsk, Moscow 108840, Russia}

\author{Enrico Landi}
\affiliation{Department of Climate and Space Sciences and Engineering, University of Michigan, Ann Arbor, MI 48109, USA}

\begin{abstract}
The identifications of \ion{Fe}{vii} emission lines in the wavelength range 193--197~\AA\ are discussed in the light of new measurements of laboratory spectra and atomic data calculations. This region is of importance to studies of solar spectra from the EUV Imaging Spectrometer (EIS) on board the Hinode spacecraft, which has its peak sensitivity at these wavelengths. Ten lines are measured, arising from seven fine structure levels in the $3p^53d^3$ configuration. Two lines  have not previously been reported and lead to new experimental energies for the $(a^2D)^3F_{2,3}$ levels. Updated experimental energies are obtained for the remaining levels. The new atomic model is used to compute theoretical values for the two density diagnostic ratios \lam196.21/\lam195.39 and \lam196.21/\lam196.06, and densities are derived from EIS spectra of coronal loop footpoints.
\end{abstract}

\section{Introduction} \label{sec:intro}

\ion{Fe}{vii} is formed at temperatures 0.2--0.5~MK in the solar atmosphere and produces many  emission lines through the extreme ultraviolet (EUV) spectrum from 160 to 300~\AA. These offer a number of temperature and density diagnostic ratios \citep{2009A&A...508..501D,2009ApJ...707..173Y}, and the ion can be used for element abundance diagnostics. For example, the ion is formed at similar temperatures to \ion{O}{v} and \ion{O}{vi}, which also have lines in the EUV, thus allowing the Fe/O abundance ratio to be measured. This is an important ratio for understanding abundance anomalies in the solar corona \citep{2014A&A...565A..14D} and solar wind \citep{2014ApJ...787..121K}. The \ion{Fe}{vii} lines also make contributions to the bandpasses of EUV spectroscopic imaging instruments such as the EUV Imaging Telescope \citep[EIT;][]{1995SoPh..162..291D}, the Transition Region and Coronal Explorer \citep[TRACE;][]{1999SoPh..187..229H}, the EUV Imager \citep[EUVI;][]{2008SSRv..136...67H}, and the Atmospheric Imaging Assembly \citep[AIA;][]{2012SoPh..275...17L}. \citet{2009ApJ...707..173Y} demonstrated that \ion{Fe}{vii} enhanced the 171 and 195~\AA\ channel response functions for TRACE by around 50\%\ at the temperature of formation of the ion, and studies of prominence eruptions have suggested EUV channels can be dominated by ion species formed at 0.2~MK \citep{2000ApJ...529..575C,2010ApJ...711...75L}.

\ion{Fe}{vii} is of particular interest for studies of active region fan loops \citep{1999SoPh..187..261S} and sunspot plumes \citep{1974ApJ...193L.143F}. The lower sections of these structures have a large emission measure at  temperatures of 0.2--0.8~MK \citep{2009ApJ...706....1L}, strongly enhancing emission lines in this region. The ion is thus valuable for measuring the temperature gradient in the loops, thus constraining the conductive energy flux.

The EUV Imaging Spectrometer \citep[EIS;][]{2007SoPh..243...19C} on the Hinode spacecraft \citep{2007SoPh..243....3K} has observed the Sun's EUV spectrum in the bandpasses 170--212 and 246-291~\AA\ for more than 14 years.  \citet{2009ApJ...706....1L}, \citet{2009ApJ...707..173Y} and \citet{2009A&A...508..501D} presented examples of  spectra from fan loops and sunspot plumes, respectively, and compared the \ion{Fe}{vii} emission with atomic models formed from the data of \citet{2008A&A...481..543W}. However, the two groups of authors came to different conclusions on line identifications in their spectra. 
In the present work we present analysis of high resolution laboratory spectra, and create a new atomic model using data from \citet{2014ApJ...788...24T} and \citet{2018MNRAS.479.1260L}. Our
focus is on \ion{Fe}{vii} lines in the region 193--197~\AA\ as this contains some of the strongest \ion{Fe}{vii} transitions and is also the region where the EIS effective area is highest. 

Section~\ref{sect.data} presents the new atomic model of \ion{Fe}{vii}, Section~\ref{sect.lab} describes the laboratory spectra, and the line identifications are discussed in Section~\ref{sect.id}. Section~\ref{sect.disc} discusses the results, particularly in relation to the \citet{2009A&A...508..501D} analyis, and Section~\ref{sect.summary} gives a summary.

\section{Atomic data and models}\label{sect.data}

A new atomic model was constructed using the dataset published by \citet{2014ApJ...788...24T}, which provided Maxwellian-averaged collision strengths (``upsilons") and radiative decay rates for 189 levels belonging to the $3p^63d^2$, $3p^53d^3$, $3p^53d4l$ ($l=s,p,d,f$) and $3p^53d5l$ ($l=s,p$) configurations. Although \citet{2014ApJ...788...24T} stated their data should be more accurate than earlier works of \citet{2005MNRAS.357..440Z} and \citet{2008A&A...481..543W}, they also commented that computational limitations meant that configuration interaction effects could not be fully modeled in the scattering calculation. This particularly affects the complex $3p^53d^3$ configuration that produces most of the EUV lines.

Only radiative decay rates for allowed transitions were provided by \citet{2014ApJ...788...24T}, so forbidden transition rates were taken from \citet{2018MNRAS.479.1260L} and, where necessary, from \citet{2008A&A...481..543W}. 

Appendix~\ref{app.data} compares radiative decay rates and upsilons between the different atomic calculations, with agreement found within 25\%\ and 30\%, respectively, except for one transition.

Comparisons of calculated $3p^63d^24p$ and $3p^53d^3$ energies with the experimental levels from \citet{ekberg81} showed deviations of $-4200$ to $+22\,600$~cm$^{-1}$ for \citet{2014ApJ...788...24T}, and $-210$ to $9570$~cm$^{-1}$ for \citet{2018MNRAS.479.1260L}. Therefore the \citet{2018MNRAS.479.1260L} energies were used for levels with no experimental energies.

Experimental energies are taken from \citet{ekberg81}, except where these were updated by \citet{2009ApJ...707..173Y}. Updated energies for five levels are given in the present work, and new energies are provided for two levels (Section~\ref{sect.id}).

The atomic data were converted to the CHIANTI\footnote{\url{https://chiantidatabase.org}.} database format. In particular, 
an assessment procedure was applied to the upsilons such that they were scaled and extrapolated using a modification of the \citet{1992A&A...254..436B} method, as described in CHIANTI Technical Report No.~4\footnote{\url{https://chiantidatabase.org/tech_reports/04_write_scups/chianti_report_04.pdf}}. The procedure identified a number of transitions for which the allowed transition upsilons did not tend towards the high-temperature limit points. This was likely due to the fact that \citet{2014ApJ...788...24T} used a more elaborate model for the structure calculation compared to the scattering calculation.  

\begin{figure*}[t]
    \centering
    \plotone{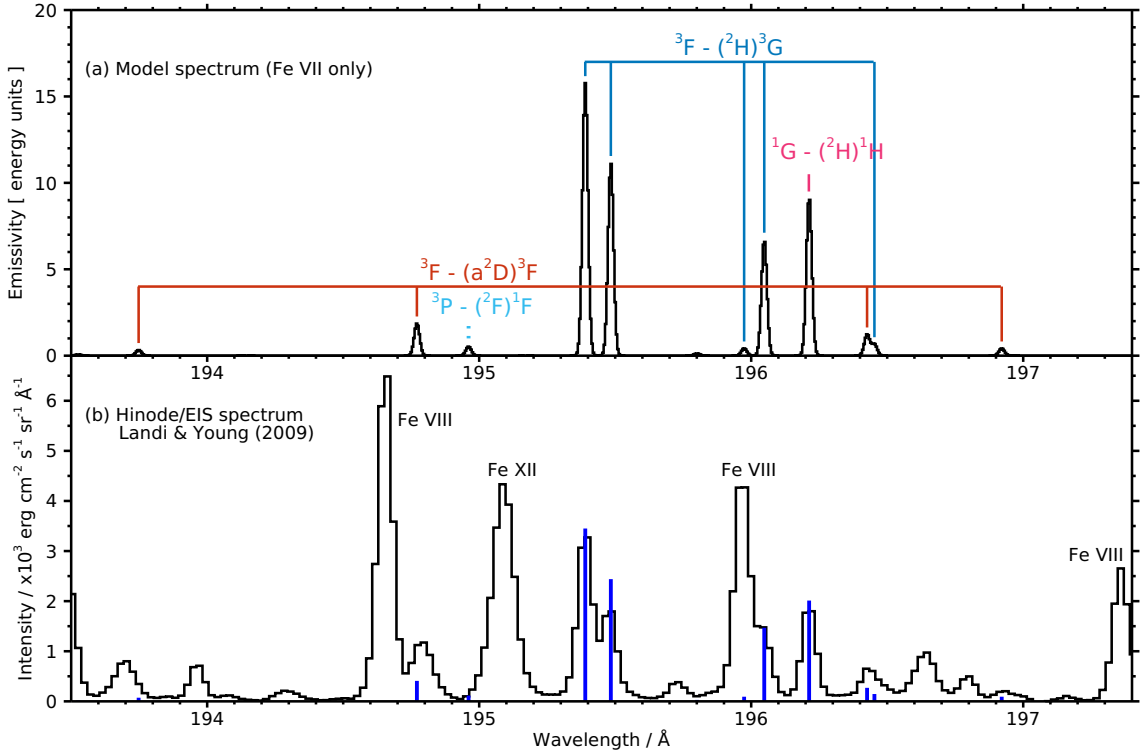}
    \caption{(a) A \ion{Fe}{vii} spectrum computed from the atomic model described in Sect.~\ref{sect.data} for a temperature of $\log\,T=5.5$ and an electron number density of $10^9$~cm$^{-3}$. The eleven lines discussed in the present work along with their multiplets are indicated. The vertical  dotted line denotes a line with a theoretical wavelength. (b) A section of the Hinode/EIS spectrum of \citet{2009ApJ...707..173Y}, with the model \ion{Fe}{vii} lines from panel (a) plotted with blue lines. Four strong lines that are not due to \ion{Fe}{vii} are identified.}
    \label{fig:spec}
\end{figure*}

With the data in the CHIANTI format, the CHIANTI IDL software was run to compute emissivities and assess the strengths of the transitions in the 193--197\AA\  region.
Although there are uncertainties in the line identifications in this region, it is clear from the atomic models that the only transitions that are relevant belong to four multiplets. In each case the transitions are decays from the $3p^53d^3$ configuration to the ground $3p^63d^2$ configuration and henceforward we will not refer to the configurations when specifying the transitions. (Thus $^3F_4$--$(^2H)^3G_5$ is shorthand for $3p^63d^2$ $^3F_4$ -- $3p^53d^3$ $(^2H)^3G_5$.)  The $3p^53d^3$ levels are assigned parent terms due to the many duplicate terms in this configuration. The parent term assignments come from the AUTOSTRUCTURE model mentioned in \citet{2009ApJ...707..173Y}. The multiplets are (in order of importance)  $^3F$--$(^2H)^3G$ (5 lines), $^1G$--$(^2H)^1H$ (1 line), $^3F$--$(a^2D)^3F$ (3 lines) and $^3P$--$(^2F)^1F$ (1 line). Figure~\ref{fig:spec}(a) shows the \ion{Fe}{vii} spectrum computed from the current atomic model for a temperature, $T$, of $10^{5.5}$~K and electron number density, $N_{\rm e}$, of $10^9$~cm$^{-3}$. The lines belonging to the four multiplets are indicated. The wavelengths of these lines are discussed in the following sections. Figure~\ref{fig:spec}(b) shows the EIS spectrum from \citet{2009ApJ...706....1L} with the model \ion{Fe}{vii} intensities from panel (a) over-plotted.

Table~\ref{tbl.em} gives the emissivities for the eleven transitions considered in this work, computed at  $\log\,T=5.5$ and  $\log\,N_{\rm e}=8,9,10$, which are values typical of the solar atmosphere. Additional values are given for $\log\,N_{\rm e}=18$, which is the approximate density of the laboratory plasma. Level indices correspond to those given in Table~2 of \citet{2014ApJ...788...24T}.
The quantity shown is:
\begin{equation}
    \epsilon_{ij} = {E_{ji} n_j A_{ji} \over N_{\rm e}} \quad\quad [{\rm erg}~{\rm cm}^{-3}~{\rm s}^{-1}]
\end{equation}
where $E_{ji}$ is the energy of the $i$--$j$ transition, $n_j$ is the population of the upper level (normalized such that $\sum_j n_j=1$), and $A_{ji}$ is the radiative decay rate. We note that the contribution function (in the format returned by the CHIANTI IDL routine \verb|gofnt|) is obtained by multiplying $\epsilon$ by  the ionization fraction and the element abundance, and dividing by $4\pi$.

Comparisons with emissivities constructed from two other atomic models---one of which is that used by \citet{2009ApJ...707..173Y}---are given in Appendix~\ref{app.em}. Agreement is found to within 26\%, except for one transition that shows differences of up to 43\%.

\begin{figure*}[t]
    \plotone{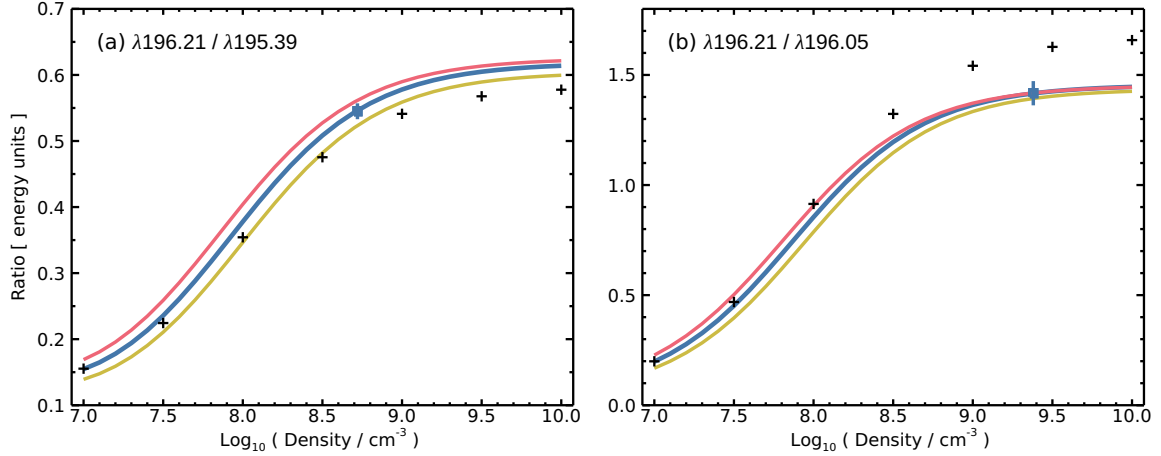}
    \caption{Predicted variation with density of the \ion{Fe}{vii} (a) \lam196.21/\lam195.39 and (b) \lam196.21/\lam196.05 ratios. The ratios are computed for temperatures $\log\,T=5.40$ (red), 5.55 (blue) and 5.70 (yellow). Black crosses show the ratios derived from the \citet{2009ApJ...707..173Y} atomic model, computed at $\log\,T=5.55$. Squares mark the ratios measured from the Hinode/EIS spectrum, with 1$\sigma$ uncertainties given as a vertical line.}
    \label{fig.dens}
\end{figure*}

\citet{2009ApJ...707..173Y} identified the \lam196.21/\lam195.39 ratio as a good density diagnostic, with sensitivity over the range $\log\,[N_{\rm e}/{\rm cm}^{-3}]=6.5$ to 9.5. From the EIS spectra they derived a density of $\log\,N_{\rm e}=8.68\pm 0.08$. However, the 195.39~\AA\ line contains a weak \ion{Fe}{x} blend \citep[noted in][]{2009ApJ...707..173Y} that was not corrected for. Adjusting for this yields a value of $9.05^{+0.20}_{-0.14}$.
The ratio curve from the new atomic model is shown in Figure~\ref{fig.dens}(a) and compared with the earlier model. It is seen that the ratio is shifted slightly to lower densities, yielding an updated density of $\log\,N_{\rm e}=8.72\pm 0.09$. 
We also show the \lam196.21/\lam196.05 ratio in Figure~\ref{fig.dens}(b) as the lines are adjacent in the spectrum. The curve is about 10\%\ lower than that from the previous model at high densities, and the measured EIS ratio is close to the high density limit giving a lower limit of $\log\,N_{\rm e}=8.99$. This value is consistent with the density of $9.02\pm 0.06$ obtained from the \ion{Mg}{vii} \lam280.72/\lam278.39, which we consider the most accurate diagnostic in this temperature range.

\begin{deluxetable*}{ccccccccccc}
\tablecaption{Laboratory measurements and emissivities for lines in the 193--197~\AA\ range.\label{tbl.em}}
\tablehead{
 &&&\multicolumn{2}{c}{Wavelength\tablenotemark{b}/\AA}&&\multicolumn{4}{c}{$\epsilon_{ij}$\tablenotemark{c}/$10^{21}$erg~cm$^{3}$~s$^{-1}$} \\
 \cline{4-5} \cline{7-10}
 \colhead{Terms} &
 \colhead{$J$--$J^\prime$} &
 \colhead{$i$--$j$\tablenotemark{a}} &
 \colhead{Present} &
 \colhead{Ekberg} &&
 \colhead{$10^8$} &
  \colhead{$10^9$}  &
  \colhead{$10^{10}$} &
    \colhead{$10^{18}$} &
  \colhead{Intensity}}
\startdata
$^3F$--$(^2H)^3G$ & 2--3 &   1--106&  196.049(6)&196.046 &&    7.64&    6.72&    6.60&    4.70&    4.55\\
& 3--3 & 2--106&  196.453(4)&196.453&&    0.76&    0.67&    0.66&    0.47&    0.67\\
& 3--4 &  2--108&  195.480(4)&\nodata&&   13.51&   11.15&   10.83&    7.15&    7.28\\
& 4--4 &  3--108&  195.971(4)&\nodata&&    0.51&    0.42&    0.41&    0.27&    2.58\tablenotemark{d}\\
& 4--5 &  3--109&  195.392(4)&195.391&&   17.23&   15.77&   15.47&   10.00&   14.25\\
\noalign{\smallskip}
$^3F$--$(a^2D)^3F$ & 2--2 &   1--110&  193.744(4)&\nodata&&    0.31&    0.33&    0.34&    0.19&    0.18\\
& 3--3 &  1--107&  194.770(4)&\nodata&&    2.01&    1.87&    1.86&    1.59&    1.06\\
& 3--4 &  2--105&  196.426(6)&196.423&&    1.27&    1.23&    1.27&    0.88&    2.61\\
& 4--4 &  3--105&  196.920(6)&196.917&&    0.43&    0.42&    0.44&    0.30&    0.36\\
\noalign{\smallskip}
$^1G$--$(^2H)^1H$ & 4--5 &  8--114&  196.216(5)&\nodata&&    6.68&    9.20&    9.56&    6.46&    6.46\tablenotemark{e}\\
\noalign{\smallskip}
$^3P$--$(^2F)^1F$ & 2--3 & 7--111&  194.961\tablenotemark{f}&\nodata&&    0.46&    0.53&    0.56&    0.31& \nodata\\
\enddata
\tablenotetext{a}{Level indices of \citet{2014ApJ...788...24T}.}
\tablenotetext{b}{Uncertainties on  the last digit of the present wavelengths are given in parentheses.}
\tablenotetext{c}{Computed at $\log\,T=5.5$ and densities $\log\,N_{\rm e}=8,9,10, 18$.}
\tablenotetext{d}{Blended with \ion{Fe}{viii}.}
\tablenotetext{e}{Laboratory intensities are normalized to this line.}
\tablenotetext{f}{Theoretical wavelength from \citet{2018MNRAS.479.1260L}.}
\end{deluxetable*}

\section{Laboratory spectra}\label{sect.lab}

Spectra in the wavelength range 90--350~\AA\ were obtained from a high resolution grazing incidence spectrograph with a 3 m (3600 lines/mm) grating. Previous spectra used for the analysis of \ion{Fe}{viii} \citep{1980OptSp..48..348R} were recorded on photographic plates, whereas Fuji Imaging Plates were used for the new spectra \citep{2017EPJWC.13203043R}. The FWHM values of the lines in the 200~\AA\ region are 0.017~\AA\ and 0.024~\AA\ for the  photoplates and  imaging plates, respectively, corresponding to spectral resolutions ($\lambda/\Delta\lambda$) of 12\,000 and 8000. (These compare with a resolution of 3000 for the Hinode/EIS instrument at 195~\AA.) Although having worse resolution, the imaging plate spectra possess a high linearity of the line intensities. The wavelengths were measured from the photoplate spectra whereas the line intensity data were obtained from the imaging plate spectra. The estimated uncertainty of the wavelengths is from 0.004~\AA\ to 0.006~\AA. The intensities are given on a relative scale without taking into account possible wavelength variation of the setup sensitivity. The spectra were excited in a vacuum spark run from a capacitor with $C = 10$ or 150~F, charged from 1 to 5~kV. The observation of the changes of the line intensities with a variation of the discharge conditions served for a separation of the lines belonging to different iron ions. Further experimental details can be found in the earlier references.

\begin{figure*}
    \centering
    \epsscale{1.0}
    \plotone{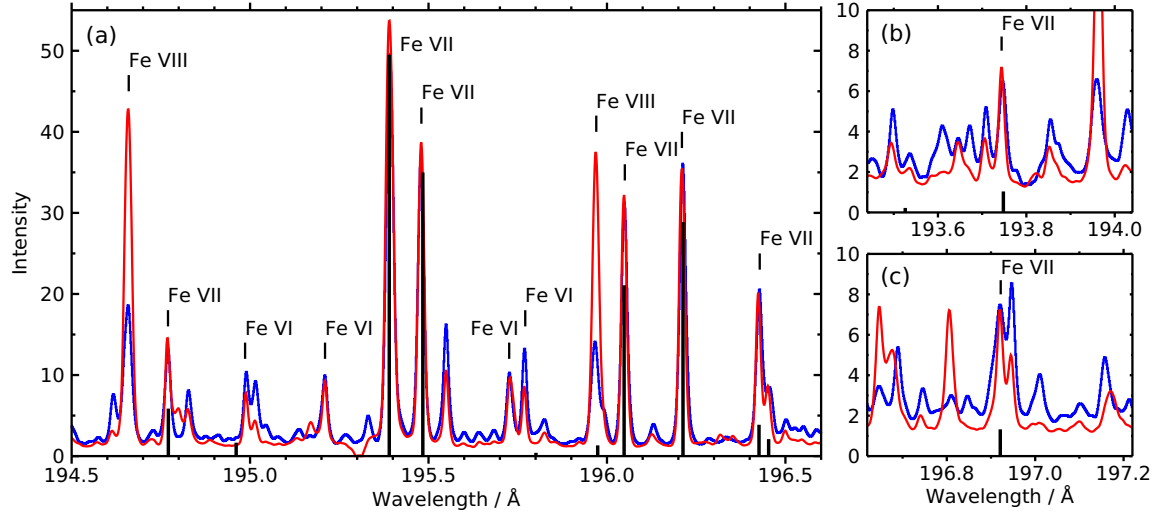}
    \caption{Two laboratory spectra obtained with the photoplates. The red and blue lines correspond to the ``hot" and ``cold" spectra, respectively. Lines due to the three iron species are indicated and distinguished by their relative intensities in the two spectra.}
    \label{fig:photo}
\end{figure*}

Figure~\ref{fig:photo} shows spectra from the photoplate data, which have a higher spectral resolution. Appendix~\ref{app.ip} discusses the imaging plate spectra, which were used for intensity measurements due to the high signal linearity.
Three sections of the iron spark spectra are shown in Figure~\ref{fig:ip}: the strong lines between 194.5 and 196.6~\AA\ (a), and the two weak lines at 193.7 and 196.9~\AA\ (b and c). In each panel the ``cold" and ``hot" spectra are shown as blue and red lines, respectively. Lines due to \ion{Fe}{viii} are about twice as strong in the hot spectra, and the strong lines at 194.66~\AA\ and 195.97~\AA\ are clearly seen. \ion{Fe}{vi} identifications are from \citet{1996PhyS...53..398A} and the lines are generally weaker in the hot spectrum. \ion{Fe}{vii} lines have comparable intensities in the two spectra and, by comparing with the model spectrum shown in Figure~\ref{fig:spec}, one can see there is a good match in terms of relative intensities. There are a number of unidentified lines, with the most significant being that at 195.53~\AA, and they are likely due to \ion{Fe}{v} or \ion{Fe}{vi}. The strong \ion{Fe}{ix} \lam197.86 line is not present, implying an absence of species hotter than \ion{Fe}{viii}.

The wavelengths measured from the laboratory spectrum are given in the third column of Table~\ref{tbl.em}, alongside the measured values of \citet{ekberg81}. The final column of Table~\ref{tbl.em} gives the line intensities measured from the imaging plate spectrum, with the values normalized to the theoretical emissivity of the line at 196.216~\AA\ at $\log\,N_{\rm e}=18$. We note that the intensities  are not corrected for the spectrograph efficiency as a function of wavelength, but this effect should be small given the narrow spectral range considered here.
The line at 195.971~\AA\ is blended with a \ion{Fe}{viii} transition that is significantly stronger, and the line at 196.922~\AA\ is blended with \ion{Fe}{vi}, although in this case the \ion{Fe}{vii} line is dominant.

\begin{deluxetable*}{cccccccccc}
\tablecaption{Experimental wavelengths.\label{tbl.wvl}}
\tablehead{
  \colhead{Terms} &
  \colhead{$J$--$J^\prime$} &
  \colhead{E81} &
  \colhead{YL09} & 
  \colhead{DZ09} &
  \colhead{Present} &&
    \colhead{Li18} 
}
\startdata
$^3F$--$(^2H)^3G$ & 2--3 & 196.046 & 196.045 & 197.358 & 196.049 &&195.737\\
  & 3--3 & 196.453 & 196.450 & 197.769 & 196.453 && 196.229 \\
  & 3--4 & 196.423 & 195.480 & 196.210 & 195.480 && 194.691\\
  & 4--4 & 196.917 & 195.970 & 196.704 & 195.971 && 195.177\\
  & 4--5 & 195.391 & 195.388 & 196.046 & 195.392 && 194.450 \\
\noalign{\smallskip}
$^3F$--$(a^2D)^3F$ & 2--2 & -- & 193.788\tablenotemark{a} & -- & 193.744 && 193.498 \\
   & 2--3 & -- & 194.728\tablenotemark{a} & -- & 194.770 && 194.453 \\
   & 3--4 & -- & 196.423 & -- & 196.426 && 196.150\\
   & 4--4 & -- & 196.918 & -- & 196.920 && 196.643\\
\noalign{\smallskip}
$^1G$--$(^2H)^1H$ & 4--5 & -- & 196.212 & 196.209 & 196.216 && 195.073\\
\enddata
\tablenotetext{a}{Estimated wavelengths (see main text).}
\end{deluxetable*}

\section{Line identifications}\label{sect.id}

Table~\ref{tbl.wvl} summarizes the experimental wavelengths in the 193--197~\AA\ range that were reported by \citet[][E81]{ekberg81}, \citet[][YL09]{2009ApJ...707..173Y} and \citet[][DZ09]{2009A&A...508..501D}. Also shown are the wavelengths obtained from the theoretical energy values of \citet[][Li18]{2018MNRAS.479.1260L}. Table~\ref{tbl.energies} compares the experimental energies with the theoretical energies of \citet[][WB08]{2008A&A...481..543W}, \citet[][TZ14]{2014ApJ...788...24T} and \citet{2018MNRAS.479.1260L}. For levels within a multiplet, the lowest energy level is listed with the energy splitting relative to this level given for the remaining levels.

\begin{deluxetable*}{cccccccccc}
\tablecaption{Energy values in cm$^{-1}$.\label{tbl.energies}}
\tablehead{
  \colhead{Term} &
  \colhead{J} &
  \colhead{Index} &
  \colhead{WB08} & 
  \colhead{TZ14} & 
  \colhead{Li18} &
  \colhead{E81} &
  \colhead{YL09} &
  \colhead{DZ09} &
  \colhead{Present} 
}
\startdata
$(^2H)^3G$ & 3 & 106 & 524\,582 & 527\,900 & 511\,940 & 510\,086 & 510\,086 & 506\,693 & 510\,080\\
           & 4 & 108 & +3082 & +2700 & +2747 & +72 & +2527 & +4016 & +2522 \\ 
           & 5 & 109 & +5313 & +4500 & +4663 & +4047 & +4047 & +5722 & +4049 \\
\noalign{\smallskip}
$(a^2D)^3F$ & 4 & 105 & 523\,308 & 527\,300 & 510\,866 &\nodata & 510\,158 & \nodata & 510\,149 \\ 
            & 3 & 107 & +3379 & +3200 & +3398 & \nodata & +3379\tablenotemark{a} & \nodata & +3274 \\
            & 2 & 110 & +5871 & +5400 & +5936 & \nodata & +5871\tablenotemark{a} & \nodata & +5986 \\
\noalign{\smallskip}
 $(^2F)^1F$ & 3 & 111 &  547\,153 & 546\,400 & 534\,202 & \nodata &\nodata& \nodata & \nodata \\
$(^2H)^1H$ & 5 & 114 & 556\,628 &552\,200 & 541\,557 & \nodata & 538\,566 & 538\,588 & 538\,578 \\ 
\enddata
\tablenotetext{a}{Estimated energies with an accuracy of $\pm$200~cm$^{-1}$.}
\end{deluxetable*}

\citet{ekberg81} was the first to perform a detailed study of the 
\ion{Fe}{vii} EUV spectrum  using laboratory spectra, and this was the starting point of the \citet{2009ApJ...707..173Y} analysis. A key discrepancy was found for the $(^2H)^3G_4$ level, which \citet{ekberg81} identified with emission lines at 196.42 and 196.92~\AA\ (Table~\ref{tbl.wvl}) leading to an energy only 72~cm$^{-1}$ greater than that of $(^2H)^3G_3$ (Table~\ref{tbl.energies}). The three theoretical works consistently yield a much larger separation of 2700--3100~cm$^{-1}$. This led to \citet{2009ApJ...707..173Y} identifying the $(^2H)^3G_{4}$ level with the strong line at 195.48~\AA, which gives  better agreement with the theoretical energy splittings. We note here an error in Table~3 of \citet{2009ApJ...707..173Y} whereby the energy of 512\,601~cm$^{-1}$ for $(^2H)^3G_{4}$ should have been given as $512\,613\pm 8$~cm$^{-1}$, which is derived from the 195.48~\AA\ line measured in the EIS spectrum. It is the latter value that is given in Table~\ref{tbl.energies}. The laboratory spectra presented here clearly demonstrate that the 195.48~\AA\ line is due to \ion{Fe}{vii}. A comparison of the model emissivities with the laboratory intensities (Table~\ref{tbl.em}) also clearly shows that the strength of the 195.48~\AA\ line is much more consistent with the $(^2H)^3G_{4}$ level than the weak line at 196.42~\AA.

With the lines measured by \citet{ekberg81} at 196.42 and 196.92~\AA\ no longer associated with $(^2H)^3G_4$, \citet{2009ApJ...707..173Y} identified them with $(a^2D)^3F_4$. Table~\ref{tbl.em} shows that the model emissivity for the 196.92~\AA\ line is in good agreement with the laboratory intensity, but the 196.42~\AA\ measured intensity is much stronger than expected. However, no other $J=4$ level identification is suitable and so we believe these identifications are correct. \citet{2009ApJ...707..173Y} did not identify the 196.92~\AA\ line in their spectrum, but they did measure a line at 196.964~\AA. Correcting this wavelength for the plasma Doppler shift gives 196.937~\AA. The intensity ratio relative to the 196.42~\AA\ line is 0.77, significantly larger than the theoretical ratio of 0.34 (Table~\ref{tbl.em}). Thus the solar line at 196.94~\AA\ appears to be blended.

\citet{2009ApJ...707..173Y} adjusted the theoretical energies of the $(a^2D)^3F_{2,3}$ levels based on the theoretical splittings relative to $(a^2D)^3F_4$ leading to improved estimates of the wavelengths of lines from these levels. It was not possible to identify these transitions in the EIS spectra, but they are apparent in the laboratory spectra close to the predicted wavelengths and are shown in Figure~\ref{fig:photo}. The emissivity comparison is good for these lines (Table~\ref{tbl.em}), giving confidence in the identifications. We therefore give new experimental energies for these two levels in Table~\ref{tbl.energies}, which we estimate to be accurate to 9~cm$^{-1}$ (Table~\ref{tbl.mix}).

\citet{ekberg81} was not able to identify the strong $^1G_4$--$(^2H)^1H_5$ transition, perhaps because no other transitions from the $(^2H)^1H_5$ level could be found in the spectrum. \citet{2009ApJ...707..173Y} identified it with the 196.21~\AA\ line as it could be clearly identified with \ion{Fe}{vii} in the EIS spectra and the intensity was consistent with nearby \ion{Fe}{vii} lines. This is confirmed with the laboratory spectra here.

The line at 196.45~\AA\ was identified by \citet{ekberg81}, but was blended in the EIS spectra. Figure~\ref{fig:photo} clearly shows this line as a ``shoulder" on the long-wavelength side of the stronger 196.42~\AA\ line and the intensity is in reasonable agreement with the model (Table~\ref{tbl.em}).

Finally we consider the $^3P_2$--$(^2F)^1F_3$ transition, which is estimated at 194.96~\AA\ using the \citet{2018MNRAS.479.1260L} theoretical energy. By comparison with the $(a^2D)^1D_2$ and $(^2H)^1H_5$ levels that are close to $(^2F)^1F_3$ and have experimental energies, we can estimate that the \citet{2018MNRAS.479.1260L} energy over-estimates the true energy by between 2300 and 3000~cm$^{-1}$. This would place the $^3P_2$--$(^2F)^1F_3$ transition between 195.84 and 196.11~\AA, close to the strong \ion{Fe}{viii} and \ion{Fe}{vii} lines in this range (Figure~\ref{fig:photo}). As no candidate can be seen, then it is possible the line is blended with one of these lines.

\section{Discussion}\label{sect.disc}

The previous section showed that the \ion{Fe}{vii} line identifications in the 193--197~\AA\ range are now complete except for the weak $^3P_2$--$(^2F)^1F_3$ transition. In this section we discuss additional issues related to these lines.

\citet{2009A&A...508..501D} presented an analysis of the EIS \ion{Fe}{vii} spectrum using an atomic model constructed from the \citet{2008A&A...481..543W} collision strengths and the radiative decay rates from his own structure calculation. He identified the observed 196.21~\AA\ line with the  $^1G_4$--$(^2H)^1H_5$ transition, consistent with \citet{2009ApJ...707..173Y}. A key difference with \citet{2009ApJ...707..173Y}, however,  is that he modified the line identifications of all of the $(^2H)^3G_J$ levels, giving the wavelengths listed in Table~\ref{tbl.wvl}. In particular, the strong $^3F_J$--$(^2H)^3G_{J+1}$  ($J=2,3,4$) transitions were matched to the observed lines at 197.36, 196.21 and 196.05~\AA. The 197.36~\AA\ line is blended with a much stronger \ion{Fe}{viii} line and the 196.21~\AA\ line is a self-blend with the $^1G_4$--$(^2H)^1H_5$ transition. These identifications were motivated by the fact that the observed line at 195.39~\AA\ seemed to be too strong in the EIS spectrum, and that the image formed in this line was more consistent with \ion{Fe}{viii}. By switching the identification to the weaker 196.05~\AA\ line these problems could be solved.

The laboratory spectrum presented here clearly demonstrates that the 195.39~\AA\ line is due to \ion{Fe}{vii}. The emissivity comparison does show that the measured line is stronger than expected compared to other lines by around 30\%. This may indicate a \ion{Fe}{vi} blend (see Appendix~\ref{app.ip}). The fact that the solar image formed from the EIS line appears to be hotter than \ion{Fe}{vii} was explained by \citet{2009ApJ...707..173Y} as due to a blend with a \ion{Fe}{x} line. 

A further problem with the \citet{2009A&A...508..501D} identifications is that they require  fortuitous blends with strong lines.  Both lines are very narrow in the solar spectra  \citep{2009ApJ...706....1L}, and thus the two \citet{2009A&A...508..501D} identifications require very precise wavelength matches, which seems an unlikely coincidence. The \citet{2009A&A...508..501D}  $(^2H)^3G_{J}$ level splittings are in worse agreement with theory compared to our new measurements (Table~\ref{tbl.energies}), and the 196.45~\AA\ line is inconsistent with the upper level being $(^2H)^3G_{5}$ as the $^3F_3$--$(^2H)^3G_{5}$ transition is forbidden. 

Although we are confident in the line identifications given in Table~\ref{tbl.wvl}, we highlight that the lines at 195.39~\AA\ and 196.42~\AA\ are significantly stronger than the other lines. Almost a factor three in the latter case. Blending is a possibility but we also note the conclusion of \citet{2014ApJ...788...24T} that their scattering calculation could not include all of the configuration interaction terms necessary to fully model the $3p^53d^3$ configuration.  We illustrate why this is important in Table~\ref{tbl.mix}, which gives the mixing coefficients for the $3p^53d^3$ levels considered here, computed with the Cowan atomic code by one of us (A.~Ryabtsev). (The listed theoretical energies come from this calculation.) All of the $(a^2D)^3F$ and $(^3H)^3G$ levels are highly mixed and so it is important that the scattering calculation accurately takes into account this mixing. We also highlight that the ``$(^2F)^1F_3$" \citep[the parent term being assigned by][]{2009ApJ...707..173Y}  level actually has a dominant contribution from $(^2D)^1F_3$ in this calculation, illustrating how mixing can change with different calculations.

\begin{deluxetable*}{cccccl}
\tablecaption{Mixing coefficients for selected \ion{Fe}{vii} levels.}\label{tbl.mix}
\tablehead{
&&\multicolumn{2}{c}{Energy/cm$^{-1}$} \\
\cline{3-4}
  \colhead{Term} &
  \colhead{$J$} & 
  \colhead{Experimental\tablenotemark{a}} &
  \colhead{Theoretical} &
  \colhead{Difference} &
  \colhead{Percentage contributions\tablenotemark{b}} 
}
\startdata
($^2H$)$^3G$ & 3  &  510080(6) & 509469 &  +611 & 44\%  $(^2H)^3G$ + 22\%  $(^2F)^3G$ + 17\%  $(^4F)^3G$ \\
            & 4 & 512602(9) & 512098 &  +504 & 48\%  $(^2H)^3G$  + 18\%  $(^4F)^3G$  + 18\%  $(^2F)^3G$ \\
            & 5 &   514129(8) & 514066 &  $+63$ & 54\%  $(^2H)^3G$  + 22\%  $(^4F)^3G$ +  12\%  $(^2F)^3G$ \\\\
\noalign{\smallskip}
($a^2D$)$^3F$ & 4 & 510149(11) & 509143 & +1006 & 37\%  $(a^2D)^3F$ + 21\%  $(^2F)^3F$ +  17\%  $(b^2D)^3F$ \\
              & 3 & 513423(9) & 512035 & +1388 & 36\%  $(a^2D)^3F$ + 19\%  $(^2F)^3F$ + 18\%  $(b^2D)^3F$ \\
              & 2 & 516135(9) & 514542 & +1593 & 43\%  $(a^2D)^3F$ + 21\%  $(b^2D)^3F$ + 17\%  $(^2F)^3F$ \\
\noalign{\smallskip}
($^2F$)$^1F$ & 3  & \nodata & 531117 & \nodata & 34\% $(^2D)^1F$ + 23\% $(^2F)^1F$  + 21\% $(a^2D)^3D$ \\
\noalign{\smallskip}
($^2H$)$^1H$  & 5 &  538578(11) & 538227 &  +351 & 92\%  $(^2H)^1H$ \\
\enddata
\tablenotetext{a}{The number in parentheses gives the uncertainty on the last digit.}
\tablenotetext{b}{All terms belong to the $3p^53d^3$ configuration; only contributions of $\ge 10$\%\ are shown.}
\end{deluxetable*}

\section{Summary}\label{sect.summary}

A new atomic model constructed from the atomic data of \cite{2014ApJ...788...24T} has been presented along with high-resolution laboratory spectra of \ion{Fe}{vii} lines. The spectra are important for confirming \ion{Fe}{vii} line identifications in the range 193--197~\AA, which is relevant to the study of spectra from the  Hinode/EIS instrument. In particular the strong lines at 195.39 and 195.48~\AA\ are clearly seen to be due to \ion{Fe}{vii}. Two new line identifications have been made, leading to new experimental energies for the 
 $(a^2D)^3F_{2,3}$ levels. The new model was recently released to the community through version 10 of the CHIANTI database \citep{2020arXiv201105211D}.

The intensities of the lines at 195.39~\AA\ and 196.42~\AA\ are stronger in the laboratory spectrum compared to the atomic model. This may be due to blending and/or uncertainties in the atomic data.

A detailed study of the entire \ion{Fe}{vii} laboratory spectrum will follow, and will be used to address some of the other issues highlighted by \citet{2009A&A...508..501D} and \citet{2009ApJ...707..173Y}.

\acknowledgements
PRY acknowledges support from the NASA Heliophysics Data Environment Enhancements program and the Hinode project.

\bibliography{sample63}{}
\bibliographystyle{aasjournal}

\newpage

\appendix

\section{Laboratory spectra obtained with imaging plates}\label{app.ip}

\begin{figure}[t]
    \centering
    \epsscale{1.0}
    \plotone{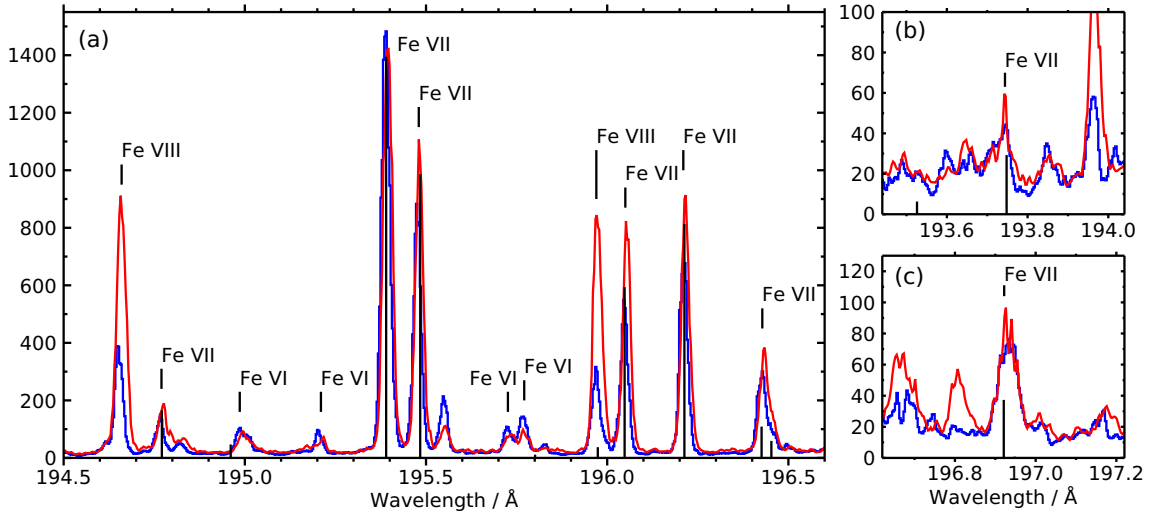}
    \caption{Two laboratory spectra obtained with the imaging plates. The red and blue lines correspond to the ``hot" and ``cold" spectra, respectively. Lines due to three iron species are indicated and distinguished by their relative intensities in the two spectra.}
    \label{fig:ip}
\end{figure}

The spectra obtained with the Fuji Imaging Plates are shown in Figure~\ref{fig:ip}. As with the photoplate spectra shown in Figure~\ref{fig:photo}, the blue and red lines correspond to ``cold" and ``hot" spectra. The latter displays stronger lines from \ion{Fe}{viii}, while the former have stronger lines from \ion{Fe}{vi}. The spectral resolution is a little worse in these spectra, as can been seen by comparing the feature at 196.4~\AA\ with Figure~\ref{fig:photo}, where two components are clearly seen. The imaging plate spectra have an almost linear intensity scaling and so were used to derive the intensities given in Table~\ref{tbl.em}. The \ion{Fe}{vii} lines are generally stronger in the red spectrum, although the strongest line at 195.39~\AA\ is a little weaker in the red spectrum. This may indicate a partial blend with a \ion{Fe}{vi} line, although \ion{Fe}{vii} dominates. This effect was not seen in the photoplate spectra (Figure~\ref{fig:photo}).

\section{Emissivity comparison}\label{app.em}

The atomic model used in this paper was largely based on the \citet{2014ApJ...788...24T} atomic data. In Table~\ref{tbl.emiss} we compare the emissivities from this model with those from two other models that are indicated with ``WB08" and ``Li08". The former is the same model as used by \citet{2009ApJ...707..173Y}, while the latter is the WB08 model but with the radiative decay rates replaced by those of \citet{2018MNRAS.479.1260L}, where available. Note that \citet{2008A&A...481..543W} and \citet{2014ApJ...788...24T} provided data for the same 189 fine structure levels, but \citet{2018MNRAS.479.1260L} gave decay rates for a reduced set of 134 levels.

The WB08 and TZ14 emissivities show good agreement, with the largest difference for the 194.77~\AA\ line, which is up to 43\%\ larger in the WB08 model. The median difference between the two models is 7\%. The effect of using the \citet{2018MNRAS.479.1260L} decay rates for the WB08 model is relatively small, with a maximum difference of 19\%\ for the weak line at 194.96~\AA. The median difference for all transitions is 2.4\%.

\begin{deluxetable}{cccccccccccc}
\tablecaption{Emissivity comparison.\label{tbl.emiss}}
\tablehead{
 &\multicolumn{3}{c}{$10^8$~cm$^{-3}$} 
 &&\multicolumn{3}{c}{$10^9$~cm$^{-3}$} 
 &&\multicolumn{3}{c}{$10^{10}$~cm$^{-3}$} \\
 \cline{2-4} \cline{6-8} \cline{10-12}
 \colhead{$\lambda$/\AA} &
 \colhead{WB08} &
  \colhead{Li18}  &
  \colhead{TZ14} &&
 \colhead{WB08} &
  \colhead{Li18}  &
  \colhead{TZ14} &&
 \colhead{WB08} &
  \colhead{Li18}  &
  \colhead{TZ14} }
\startdata
  196.05&    6.67&    6.49&    7.65&&    5.64&    5.73&    6.72&&    5.49&    5.60&    6.60\\
  196.46&    0.71&    0.63&    0.76&&    0.60&    0.55&    0.67&&    0.58&    0.54&    0.66\\
  195.48&   13.19&   12.88&   13.51&&   10.89&   10.89&   11.15&&   10.55&   10.60&   10.83\\
  195.97&    0.49&    0.47&    0.51&&    0.40&    0.39&    0.42&&    0.39&    0.38&    0.41\\
  195.39&   17.06&   16.87&   17.22&&   15.86&   15.65&   15.77&&   15.55&   15.36&   15.46\\
   \noalign{\smallskip}
  193.75&    0.38&    0.37&    0.31&&    0.41&    0.41&    0.33&&    0.43&    0.43&    0.34\\
  194.77&    2.87&    2.76&    2.01&&    2.56&    2.55&    1.87&&    2.54&    2.54&    1.86\\
  196.42&    1.00&    1.06&    1.27&&    1.09&    1.17&    1.23&&    1.20&    1.29&    1.27\\
  196.92&    0.44&    0.38&    0.43&&    0.48&    0.41&    0.42&&    0.53&    0.46&    0.44\\
   \noalign{\smallskip}
 196.21&    6.18&    6.31&    6.67&&    8.64&    8.65&    9.20&&    9.02&    8.99&    9.56\\
   \noalign{\smallskip}
  194.96&    0.45&    0.37&    0.46&&    0.54&    0.44&    0.53&&    0.57&    0.46&    0.56\\
\enddata
\end{deluxetable}

\section{Comparison of decay rates and upsilons}\label{app.data}

Appendix~\ref{app.em} compared emissivities computed from three atomic models. Here we compare radiative decay rates and upsilons for the strongest transitions. 
Tables~\ref{tbl.aval} gives decay rates from WB08, TZ14 and Li18, and also those from \citet[DZ09]{2009A&A...508..501D}. For the decay rates the agreement is generally very good. The TZ14 rates are larger than the other two calculations for all but two of the transitions. The biggest discrepancy is for the $^3F_3$--$(a^2D)^3F_4$ transition, which is 92\%\ higher than WB08 and 65\%\ higher than Li18. The WB08 and Li18 data-sets agree very well for the $^2F$--$(^2H)^3G$ transitions, with decay rates within 6\%. For the other two multiplets agreement is within 25\%. A problem can be seen with the DZ09 rates whereby the $^2F$--$(^2H)^3G$ 3--4 transition is weaker than the 2--3 transition. This may be a tabulation error.

\begin{deluxetable}{cccccccc}
\tablecaption{Radiative decay rate comparison.\label{tbl.aval}}
\tablehead{
  \colhead{Term} &
  \colhead{$J$--$J^\prime$} &
  \colhead{WB08} &
  \colhead{DZ09} & 
  \colhead{TZ14} & 
  \colhead{Li18} 
}
\startdata
$^3F$--$(^2H)^3G$ & 2--3 & 2.86(+10) & 3.30(+10) & 3.82(+10) & 3.02(+10)\\
  & 3--3 & 3.05(+9) &\nodata & 3.81(+9) & 2.91(+9) \\
  & 3--4 & 4.27(+10) & 2.40(+10) & 4.90(+10) & 4.14(+10) \\
  & 4--4 & 1.59(+9) & \nodata & 1.86(+9) & 1.50(+9) \\
  & 4--5 & 5.03(+10) & 4.60(+10) & 5.78(+10) & 4.93(+10)\\
\noalign{\smallskip}
$^3F$--$(a^2D)^3F$ & 2--2 & 1.12(+9) & \nodata & 1.19(+9) & 1.07(+9) \\
   & 2--3 & 1.12(+10) & \nodata & 9.00(+9) & 8.44(+9) \\
   & 3--4 & 1.76(+9) &\nodata & 3.38(+9) & 2.05(+9) \\
   & 4--4 & 7.77(+8) &\nodata & 1.16(+9) & 7.28(+8)\\
\noalign{\smallskip}
$^1G$--$(^2H)^1H$ & 4--5 & 3.80(+10) & 3.50(+10) &4.04(+10) & 4.12(+10) \\
\noalign{\smallskip}
$^3P$--$(^2H)^1F$ & 2--3 & 1.31(+9) & \nodata & 1.58(+9) & 1.32(+9) \\
\enddata
\end{deluxetable}

For each of the eight $3p^53d^3$ emitting levels considered in this work we compare the WB08 and TZ14 upsilons for the dominant transitions to these levels in Table~\ref{tbl.ups}. As expected, the differences mirror those seen in the decay rate calculation. The largest difference is for $^3F_3$--$(a^2D)^3F_4$, for which the TZ14 upsilon is 72\%\ higher than the WB08 upsilon. The upsilons for other transitions are within 30\%.

\begin{deluxetable}{cccc}
\tablecaption{Comparison of upsilons calculated at $\log\,T=5.50$.\label{tbl.ups}}
\tablehead{
  \colhead{Upper} &
  \colhead{Lower} &
  \colhead{WB08} &
  \colhead{TZ14} 
}
\startdata
$(^2H)^3G_5$ & $^3F_4$ & 3.49 & 3.77 \\
$(^2H)^3G_4$ & $^3F_3$ & 2.32 & 2.54 \\
$(^2H)^3G_3$ & $^3F_2$ & 1.22 & 1.58 \\
\noalign{\smallskip}
$(a^2D)^3F_4$ & $^3F_3$ & 0.110 & 0.189 \\
  & $^3F_4$ & 0.070 & 0.078 \\
$(a^2D)^3F_3$ & $^3F_2$ & 0.484 & 0.366 \\
$(a^2D)^3F_2$& $^3F_2$ & 0.035 & 0.035 \\
\noalign{\smallskip}
$(^2H)^1H_5$ & $^1G_4$ & 2.85 & 3.02 \\
\noalign{\smallskip}
$(^2H)^1F_3$ & $^3P_2$ & 0.077 & 0.092\\
\enddata
\end{deluxetable}

\end{document}